\documentclass[a4paper,fleqn]{cas-dc}
\usepackage{setspace}
\usepackage[numbers]{natbib}
\usepackage{caption}
\usepackage{subcaption}
\usepackage{csquotes}
\usepackage{float}

\def\tsc#1{\csdef{#1}{\textsc{\lowercase{#1}}\xspace}}
\tsc{WGM}
\tsc{QE}
\tsc{EP}
\tsc{PMS}
\tsc{BEC}
\tsc{DE}

\begin{document}

\let\WriteBookmarks\relax
\def\floatpagepagefraction{1}
\def\textpagefraction{.001}

\shorttitle{Joint-Stream Malignant Region learning}

\shortauthors{CV Abdul Rehman et~al.}

\title [mode = title]{Joint Stream: Malignant Region Learning for Breast Cancer Diagnosis} 



%
\author[1]{Abdul Rehman} 



\ead{msds20083@itu.edu.pk}



\affiliation[1]{organization={Intelligent Machine Lab},
 addressline={Information Technology University}, 
 city={Lahore},
 postcode={54000}, 
 country={Pakistan}}

\author[2]{Sarfaraz Hussein}
\author[1]{Waqas Sultani}
\cormark[1]
\ead{waqas.sultani@itu.edu.pk}


\affiliation[2]{organization={Machine Learning and Data Science @ The Home Depot},
 country={USA}}



\cortext[cor1]{Corresponding author}



\begin{abstract}
Early diagnosis of breast cancer (BC) significantly contributes to reducing the mortality rate worldwide. The detection of different factors and biomarkers such as Estrogen receptor (ER), Progesterone receptor (PR), Human epidermal growth factor receptor 2 (HER2) gene, Histological grade (HG), Auxiliary lymph node (ALN) status, and Molecular subtype (MS) can play a significant role in improved BC diagnosis.  However, the existing methods predict only a single factor which makes them less suitable to use in diagnosis and designing a strategy for treatment.

In this paper, we propose to classify the six essential indicating factors (ER, PR, HER2, ALN, HG, MS) for early BC diagnosis using H\&E stained WSI's. 
To precisely capture local neighboring relationships,  we use spatial and frequency domain information from the large patch size of WSI's malignant regions. Furthermore, to cater
the variable number of regions of interest sizes and give due attention to each region, we propose a malignant region learning attention network.  Our experimental results demonstrate that combining spatial and frequency information using the malignant region learning module significantly improves multi-factor and single-factor classification performance on publicly available datasets.

\end{abstract}



\maketitle
\section{Introduction}
\label{section:intro}

According to the World Health Organization (WHO), breast cancer is among the top five most common types of cancers worldwide \cite{ferlay2021cancer}. Only in 2020, almost 264,121 new cases of female BC were reported, and 42,280 women died of this cancer in the United States. The treatment at the early stages of BC can be highly effective, and the survival probability is more than  90\%.  Thanks to early detection and treatment improvements  \cite{nci_2022}, the death rate decreased by 42\% from 1989 to 2019.


For the diagnosis and treatment of BC, a biopsy has been the gold standard, and so far, the most reliable and definitive way.{
The early identification of the naturally occurring hormone types (positive or negative), estrogen receptor (ER), progesterone receptor (PR), and human epidermal growth factor receptor 2 (HER2) gene provides the idea about the best way to treat or prevent cancer from recurring \cite{onitilo2009breast}.  
The immunohistochemical determination of breast cancer subtypes concerning ER, PR, and HER2 status can contribute to improved selection of therapy and patient care \cite{effi2016breast}. For example, patients with negative labels for all ER, PR, and HER2 have much less chance of survival as compared to the patient in which one of them is positive \cite{foulkes2010triple}. Therefore,  employing the positive and negative labels of ER, PR, and HER2, we can decide on a better strategy for treatment \cite{inic2014difference}.
The other two factors, histological grade (HG), and auxiliary lymph node (ALN) status are also widely accepted and powerful indicators of the BC prognosis \cite{rakha2010breast,choi2017preoperative}.
HG is based on the degree of differentiation of the tumor tissue and for the grade assessment, trained pathologists use H\&E (Hematoxylin and Eosin) stained tumor tissue slides with standard protocol \cite{rakha2010breast}. ALN status is the most important prognostic factor for overall BC survival and is evaluated before the surgery through multi-model imaging, physical examination, and mammography \cite{choi2017preoperative}.   Finally, Molecular subtypes (MS) of invasive breast cancer are based on genes that control how cancer cells behave and the information is used for developing the treatment.
\begin{figure*}[ht]
\centering
\includegraphics[width=0.98\textwidth]{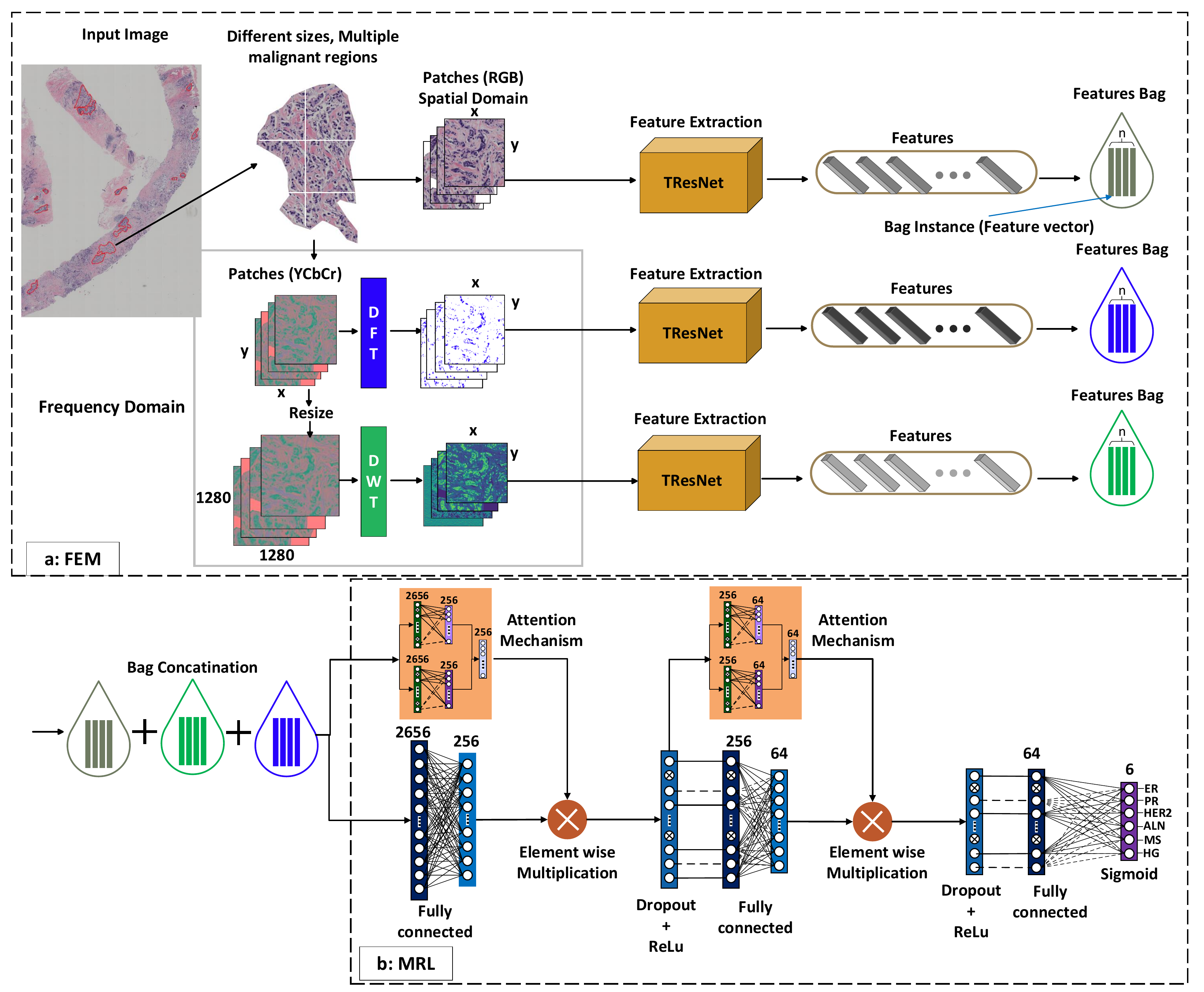}
\caption{The pipeline of the proposed joint stream: malignant region learning for multi-label classification. After extracting patches of size ($x \times y$) from the malignant regions, we extract the spatial and frequency domain features using the TresNet-xl \cite{ridnik2021asymmetric}. For the frequency domain features, we convert the patch's color space from RGB (Red, Green, Blue) to YCbCr (Luminance, Chroma: Blue, Chroma: Red), compute 
Discrete Fourier transform (DFT) and Discrete wavelet transform (DWT). Finally, we combine the spatial and frequency feature bags using the malignant region learning (MRL) module to obtain the multi-label prediction. Where {A} represents the output of the attention mechanism.}
\label{fig:pipeline}
\end{figure*}


{In the diagnostic workflow, a pathologist visually analyzes the stained tissue slides using the microscope. However, due to the recent development in imaging technology, a digital scanner takes a digital whole slide image (WSI) where the pathologists can analyze the image manually and with the help of computer vision algorithms \cite{farahani2015whole}. The WSI of traditional H\&E stained tissues is taken from the biopsy sample.  A WSI generally contains two types of regions (malignant and non-malignant) and in the malignant regions of interest (ROI) the classification scale depends on the nuclear formation, shape, and intensity of the cell division. 
It contains a lot of information (takes up to 8GB) because of the higher magnification level with more than 100 thousand pixels in both dimensions. Given the WSI image, using machine-assisted deep learning models, it is possible to determine the status of BC biomarkers and prognostic factors   (ER, PR, HER2, ALN, HG, MS) \cite{esteva2004prognostic} using the WSI of H\&E stained tissues only }\cite{naik2020deep, shaban2020context, xu2021predicting}.

{Recently there has been growing interest in using deep convolutional neural networks (CNN) for the classification of BC-related factors and biomarkers employing WSI images.  {This includes
grading classification {\cite{jimenez2019deep,koohababni2018nuclei,shaban2020context,wang2022classification}, ALN status classification \cite{xu2021predicting}, {HER2 overexpression estimation \cite{borquez2023uncertainty} }  and ER status classification \cite{naik2020deep}. Most of the  CNN's-based WSI's classification methods  use patch size of $299 \times 299$ or $256 \times 256$ pixels \cite{tsai2021deep,golatkar2018classification, lu2021data}. Although appealing, due to the small patch size, these methods ignore the spatial relationships between neighboring pixels. Therefore, it is desired to have larger size patches to capture the important feature information  \cite{shaban2020context}. Authors in \cite{jimenez2019deep,koohababni2018nuclei,shaban2020context,xu2021predicting,naik2020deep} classify only a single indicating biomarker or prognostic factor from the six (ER, PR, HER2, ALN, HG, MS)  for BC and \cite{gao2024transformer,zhang2023whole} classify the WSI is normal or malignant (metastasis) at a time. However, for a thorough automated diagnosis, a model predicting multiple BC contributing factors would be more useful \cite{board_2022}. 

{In practical applications, WSIs and the patches of WSI's are commonly stored as compressed data following the JPEG standard and decoded into the spatial domain (RGB) data format. For feature extraction, most of the approaches \cite{jimenez2019deep,koohababni2018nuclei,shaban2020context,xu2021predicting} use only RGB images. However,  the frequency analysis of WSI contains complementary information due to the presence of repetitive patterns of cells in WSI and potentially be able to develop improved deep-learning models for histopathology pattern recognition. 
However, instead of using a complete WSI, the malignant regions provide more discriminative information for predicting multiple factors.


The architecture of the proposed methodology is shown in Figure \ref{fig:pipeline}.  The contributions are
summarized as follows.\\
•  Classify multiple biomarkers (ER, PR, HER2) and prognosis factors (HG and ALN status) for early diagnosis by utilizing the malignant region of  H\&E stained WSI images.\\
•  Utilizing the frequency domain to capture unique characteristics of cell shape, size, and structure.
\\
•  Integrate spatial and frequency domain information to improve discriminative capabilities.\\
•  Adaptively extract patches from ROIs of varying sizes, prioritizing malignant ROIs.\\
•   Applied state-of-the-art (SOTA) methods: Multiple Instance Learning (MIL), Gated MIL [23] and propose the  Malignant Region Learning (MRL) to focus on crucial patches for carrying more discriminative and class-specific information, by giving more
attention to important patches.\\
•  To the best of our knowledge, we are
the first ones to combine both spatial and frequency information to predict the multi-factors (ER, PR, HER2, ALN, HG, MS).\\
•  Experimental results demonstrate significant improvements in multi-factor classification performance on publicly available datasets.\\

We have organized our paper as follows: section \ref{section:related_work} describes related research, section \ref{section:methodology} provides details of the proposed methodology, section \ref{section_ER} describes experimental results and analysis, and finally, section \ref{section_conclusion}, concludes the paper.

\section{Related Work}
\label{section:related_work}

{In this section, we briefly review recent work to understand the state-of-the-art methods for WSI classification.
\noindent\textbf{Spatial Analysis of WSI:}
Due to the large size of a WSI and limited computational resources, several efficient computer-assisted models have been proposed recently. Golatkar \emph{et al} \cite{golatkar2018classification} proposed a single-label multi-class CNN classifier for the H\&E stained breast tissues. 
To tackle the computational challenge, they extracted $299 \times 299$ pixels overlapped patches from the WSI and picked the high nuclear density patches for model training. They used the threshold value from the ratio of blue and red channel intensities for appropriate patch selection.
After that, they used a CNN-based pre-trained modified Inception-v3 \cite{szegedy2016rethinking} model for patch prediction. 
For image class prediction (normal tissue, benign lesion, in situ carcinoma, and invasive
carcinoma), they combined the patch-based predictions using majority voting to determine the class of the entire image. The approach was evaluated on the publicly available BreAst Cancer Histology Challenge- 2018 (BACH) dataset \cite{aresta2019bach}.\\
{ Secondly, Anisuzzaman\emph{et al} \cite{anisuzzaman2021deep} applied multiple deep learning models for WSI multiclass classification.  Finally,  they chose the VGG19 \cite{simonyan2015deep} and then modify for the experiments on Osteosarcomas \cite{arunachalam2019viable}  dataset. Due to the memory limitation, they downsampled the image size of $375 \times 375$ rather than the original image size of  $1024 \times 1024$.          }

Kamyar \emph{et al} \cite{nazeri2018two} proposed two consecutive CNN for BC histology H\&E stained tissues image classification.
The first CNN was used to extract the most salient feature maps from all patches in an image. These features are then stacked to form a spatially smaller 3D input for the second CNN. Similar to \cite{golatkar2018classification}, the approach was evaluated on the BACH-2018 dataset.

A WSI holds two types of regions: malignant and non-malignant. As compared to non-malignant regions, malignant regions hold more valuable variation information. {For the classification of the (malignant and non-malignant) histological images,
 Chattopadhyay \emph{et al} \cite{chattopadhyay2022mtrre} proposed a CNN-based model and used the BreakHis image dataset \cite{spanhol2015dataset} to perform the experiments. In their proposed model, they introduced a contrasting approach of dual residual block combined with the recurrent network to overcome the vanishing gradient problem. Similarly, for the BreakHis image dataset, Hayder A. Khikani \emph{et al}  \cite{khikani2022breast} proposed a capsule-based CNN model  for the classification of the malignant and non-malignant) histological images. Due to the small convolutional kernels and the Res2Net block, their proposed model also had fewer parameters.   }

 After the sentinel node biopsy, a pathologist determines the ALN status. Recently, Feng Xu \emph{et al} \cite{xu2021predicting} introduced a new large-scale multi-factor (ER, PR, HER2, ALN, HG, MS) dataset named: Breast Cancer Core Needle Biopsy (BNCB) dataset. In their studies, the authors proposed an instance attention method to classify only ALN status. 
 The framework first extracted the annotated malignant region from the WSI and then extracted the non-overlapped patch size ($228 \times 228$) from the extracted malignant region.  
These patches are divided into bags, where each bag contains a fixed number of patches. Multiple instances learning MIL \cite{ilse2018attention} attention module to used to aggregate features within each bag. Finally, they proposed two methods with or without clinical information features. 

Due to the unavailability of annotated WSI's, Ming Y.Lu \emph{et al} \cite{lu2021data} proposed instance attention aggregation-based weakly supervised detection framework 
using H\&E stained tissues WSI for the 
high diagnostic valuable region identification and classifier for multi-class classification of sub-regions. 
After removing the background region from each image and extracting the patch of size $256 \times 256$ pixels, they used a pre-trained ResNet50 on the ImageNet dataset for feature extraction and the Gated-MIL attention \cite{ilse2018attention} aggregation mechanism for indicating the valuable instances and their classification. Based on extracted features, the trained attention network is used for picking the valuable patches based on the attention score. For each class and valuable region identification, they trained a separate neural network binary classifier.\\ 
{Similarly, for the Camelyon16 \cite{bejnordi2017diagnostic} dataset, Zhang \emph{et al} \cite{zhang2023whole} proposed a three-stage classification framework for whole slide beast cancer pathology images. They pre-process the data in Stage 1 and in Stage 2 they used two deep-learning models and Swin-transformers for patch-based classification and valuable cancer region heatmap generation. In the final stage, using the medial statistical 
features extracted from the lesion region heatmaps of the whole slide, and 
then a SVM-based classifier is trained for the classification of the whole slide images. }

To deal with the computational challenge of a complex histopathjalogy dataset,\cite{nazeri2018two, zhang2023whole,lu2021data} first extract patches from the WSI and then extract features from the patches by  using pre-trained models for the most salient and more similar features from the similar patches. Few other methods, high nuclear density patches \cite{golatkar2018classification} or downsampling \cite{anisuzzaman2021deep}, which leverages the learned representations from pre-trained models. Consequently, feature extraction via pretrained models methods \cite{nazeri2018two,sangle2023accumulated,zhang2023whole,lu2021data}   become  a crucial component in handling large and complex datasets. 


\noindent\textbf{Frequency Analysis of WSI:}
The spatial information of a WSI is mostly used for both supervised and weakly supervised learning, however, the use of the frequency domain along with the spatial domain is not thoroughly investigated, especially for WSI analysis. For natural image classification, Kai Xu \emph{et al} \cite{xu2020learning} used the information from the frequency domain to perform the classification tasks. W Luo \emph{et al} \cite{luo2021frequency} proposed a  light-weight CNN-based architecture for the WSI segmentation  using the frequency information and the proposed model reduced the  bandwidth requirement for CPU-GPU transmission by reduction of 96\% parameters and the floating-point operations 
by 98\% as compared to the common CNN-based method with spatial information.  
Abdullah-Al Nahid \emph{et al} \cite{nahid2018histopathological} proposed a CNN for the image size {$700 \times 460$} and learned from the hand-crafted and frequency information for the Histopathological Breast-Image (Benign and Malignant) classification. 
For spatial feature extraction, they used Contourlet Transform (CT), Histogram information, and Local Binary Pattern (LBP), and for frequency analysis, the author used the Discrete Fourier Transform (DFT) and the Discrete Cosine Transform (DCT). The experiments were performed using the BreakHis image dataset \cite{spanhol2015dataset}}. However, the authors have not performed a joint analysis of frequency and spatial domain and dealt with a single-factor binary class problem. Although the combination of multiple sources has shown improved performance compared to any single domain features \cite{sangle2021covid,sangle2024covid}.
 In addition to the above-mentioned approaches, the previous approaches for WSI classification are based on feature attention learning. \cite{hou2016patch, xu2015deep, chikontwe2020multiple, lu2021data}.  The attention methods are divided into two types: Instance-based and embedding-based approaches.} These approaches include binary class \cite{xu2021predicting, naik2020deep} or multi-class \cite{nazeri2018two, shaban2020context, shao2021transmil, jimenez2019deep} frameworks. 

In contrast to the research works mentioned earlier, in our paper, we classify the six essential BC indicates bio-markers and factors (ER, PR, HER2, ALN, HG, MS) by explicitly combining spatial and frequency information { without using any clinical information for early BC diagnosis. Instead of choosing a fixed number of patches from the WSI, we tackle a variable number of malignant ROIs from the image and use the MRL mechanism for effective patch integration to improve the classification.}

\begin{figure*}[ht]
\centering
\includegraphics[width=0.98\textwidth]{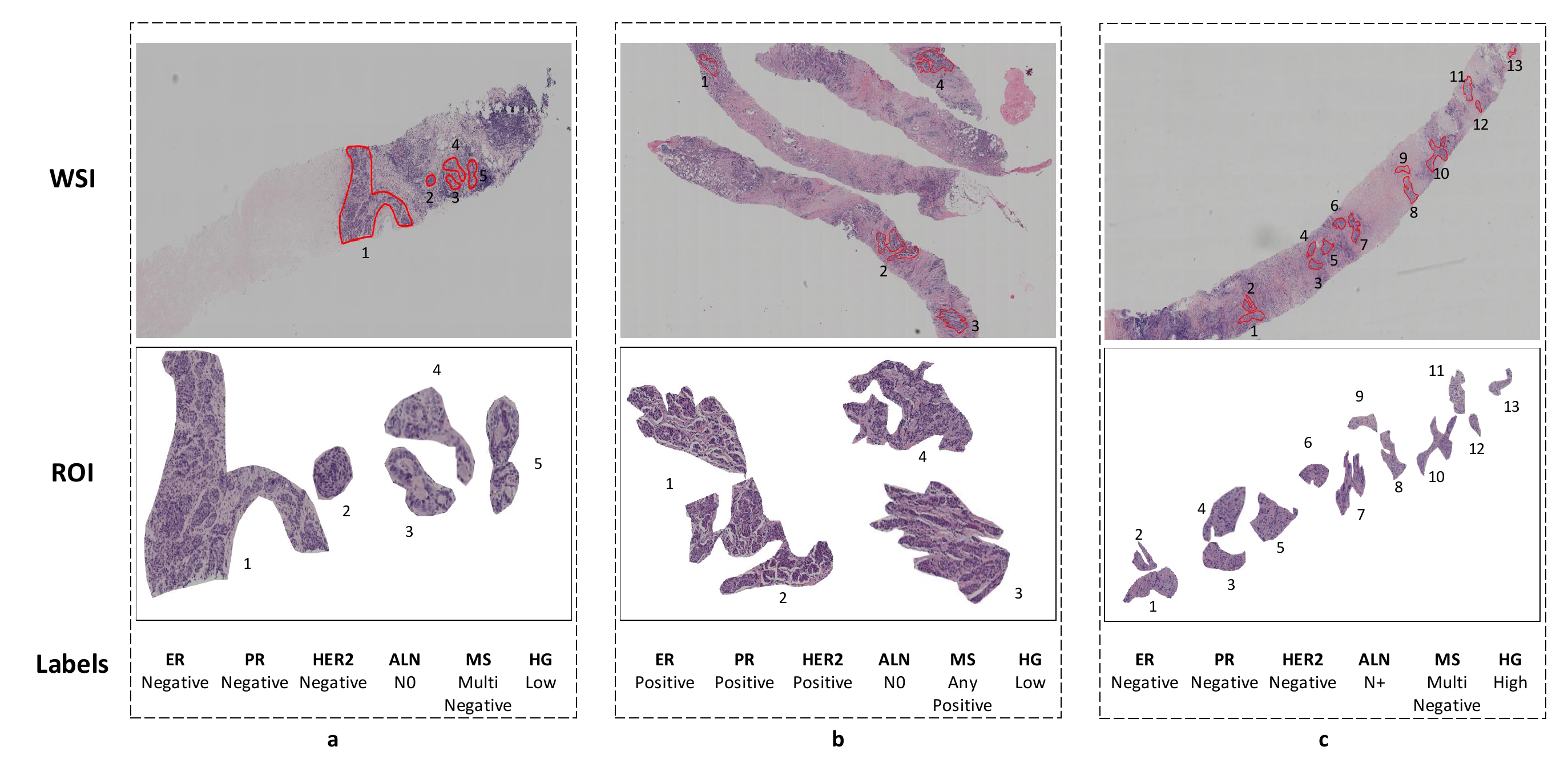}
\caption{Examples of different WSIs and their malignant ROIs with class labels from Breast Cancer Core Needle Biopsy (BNCB) dataset: The shapes, sizes, and numbers of malignant ROIs from all the WSIs are different. The class of Estrogen receptor (ER),
Progesterone receptor (PR), Human epidermal growth factor receptor 2 (HER2) gene, Histological grade (HG), Auxiliary lymph
node (ALN) status, and Molecular subtype (MS) are also given.}
\label{fig:dataset_show}
\end{figure*}

\section{Methodology}
\label{section:methodology}

Our goal is to classify the multiple BC indicating factors: ER, PR, HER2, ALN status, MS, and HG from  H\&E stained core needle biopsy WSIs. Our key observations are: 1) For the diagnosis and treatment of BC, the detection of only one indicating single factor is not enough and the method should output multiple indicating factors for robust BC detection; 2) Repetition of different patterns (nuclei/cells shape, size, and structure) are captured well in the frequency domain and therefore frequency information should be used along the spatial analysis; 3) All malignant regions are not equally important and intelligently integrating information from different patches should be helpful for more accurate prediction; 4) Since different images contain different sizes of malignant regions, therefore,  the method should be able to handle a variable number of malignant patches. Below, we provide the details of each component of the proposed approach.

\subsection{Patch Extraction}
\label{section:ROI}
A WSI is generally based on two types of regions: malignant and non-malignant. Malignant regions have irregular shapes and different sizes and their annotations need expert pathologists with several years of experience. Fortunately, the WSI datasets 1) Breast Cancer Core Needle Biopsy (BNCB) \cite{xu2021predicting} and 2) BACH \cite{aresta2019bach} have annotations available with them and we use those annotations during training. Some of the ROIs for the BNCB dataset are shown in Figure \ref{fig:dataset_show}.

\label{section:patch_extraction}

{Given the ROIs from all WSIs, we extract $640 \times 640$ overlapping patches. Note that the overlapped patches of large sizes ($640 \times 640$ vs $224 \times 224$) ensure that they hold better spatial neighboring information \cite{golatkar2018classification, nazeri2018two}.}
From the  WSI  of the average size of 11486 $\times$ 31000 pixels, we use the sliding window of half of the patch size for overlapping.   
We ignore the small ROIs (less than the patch sizes) and patches with 
a blank ratio greater than 0.3. From the BNCB training, validation, and test cohorts, we got the 15357, 4071, and 4720 patches respectively.

\subsection{Frequency Domain Transformations}
\label{section:freq_domain}

\begin{figure}[t]
\centering
\includegraphics[width=0.5\textwidth]{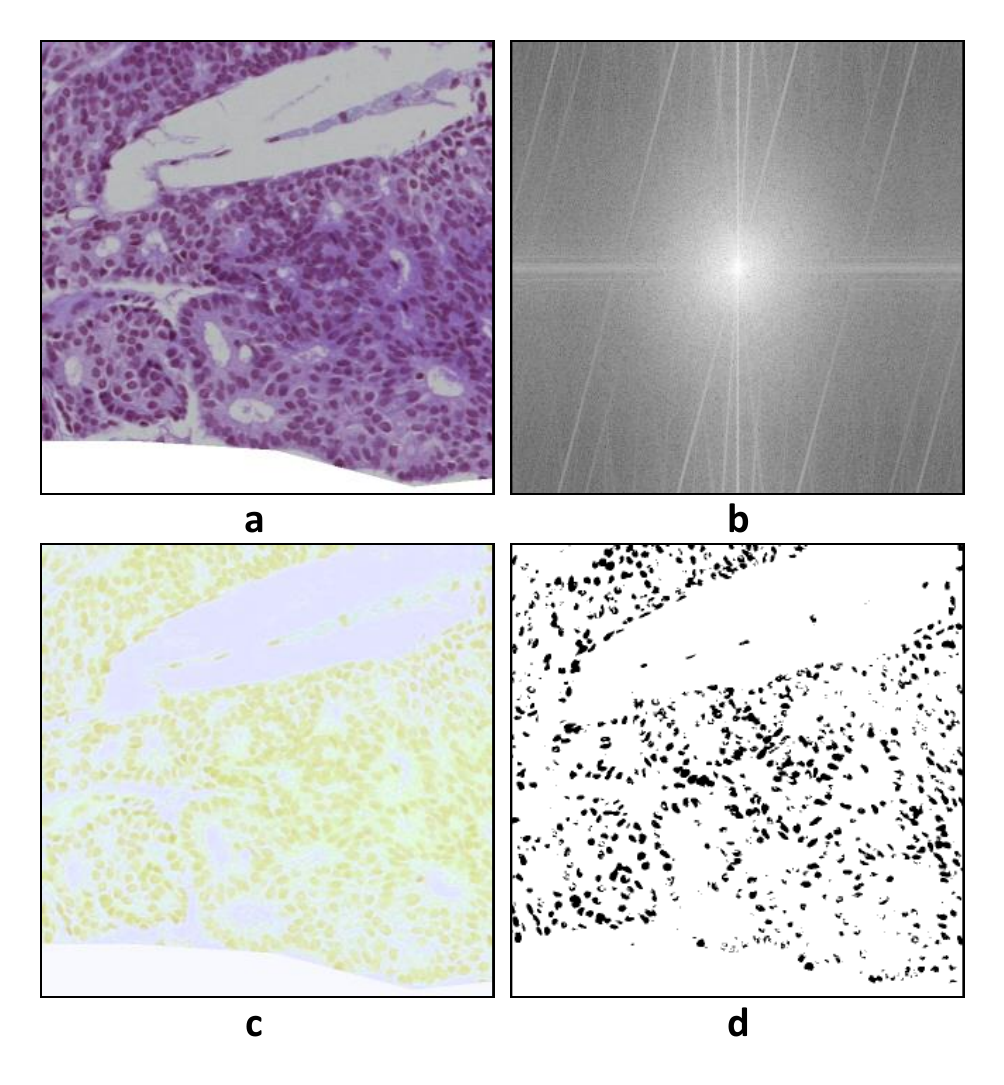}
\caption{The conversion of the spatial to frequency domain (DFT). a) original image in the spatial domain, b) single channel grayscale fourier frequency spectrum c) the real value of three channels of DFT, and d)  thresholded image where the mean value of (c) is used as a threshold. The figures show that images (c) and (d) hold very specific nuclear density information. }
\label{fig:dft_comerison}
\end{figure}

Before transforming the spatial to the frequency domain, we convert the image space from RGB to YCbCr because of the superior performance of YCbCr in frequency domain \cite{yousaf2022fake}, image enhancement and compression \cite{gopinathan2016study,midha2014analysis}, and effective recognition \cite{gede2018face}. For RGB to YCbCr space conversion, similar to  \cite{yousaf2022fake,luo2021frequency}, we use the following conversion formula.
\begin{eqnarray}\label{1} 
\begin{split}
Y&= (\alpha \times R) + (\beta \times G)+(\gamma \times B),\\
Cr&=(R-Y)\times0.713,\\
Cb&= (B-Y)\times0.564,\\
\text{where}&\\
&\alpha+\beta+\gamma=1,\\
&\alpha=0.299, \phantom{a}\beta=0.587,\phantom{a}\gamma=0.114.
\end{split}
\end{eqnarray}
To capture the frequency information from a YCbCr patch, we compute discrete fourier transform (DFT) and discrete wavelet transform (DWT) for each patch of the WSI. DWT decomposes spatial channel information into four subbands and DFT contains discriminative information which fully describes the spatial domain image.

\noindent\textbf{a) DWT :}
Before the DWT transformation, we resize the patch size from ($640\times640$) to ($1280\times1280$); because DWT decomposes
an image into four (High-high (HH), High-low (HL), Low-high (LH),
and Low-low (LL)) subbands images.  Each sub-band is down-sampled by a factor
of two in both vertical and horizontal directions.

HH region preserves high-frequency components in
both horizontal and vertical directions, LH preserves
low-frequency components in the vertical direction and high-frequency
components in the horizontal direction, HL preserves the high-frequency
components in the horizontal direction and low-frequency
components in the vertical direction and, finally,
LL preserves low-frequency components in both horizontal and vertical directions \cite{yousaf2022fake}. 
We concatenate the sub-bands and use the following structure for the final $640\times640\times12$ output.
\begin{eqnarray}\label{3}
\begin{split}
dw_{0}= &D^{y},\hspace{1cm} dw_{1}= D^{Cb},\hspace{1cm} dw_{2}= D^{Cr}, \\
{\mathcal{D}}'=& (dw_{0}[0],dw_{0}[1,0],dw_{0}[1,1],dw_{0}[1,2],\\& dw_{1}[0],dw_{1}[1,0],dw_{1}[1,1],dw_{1}[1,2],\\& dw_{2}[0],dw_{2}[1,0],dw_{2}[1,1],dw_{2}[1,2]).\\
\end{split}
\end{eqnarray}
Where ${D}^i$ represent the  Discrete Wavelet Transform function and
${dw}_n$ contains the big array of subband coefficients. Finally 
the ${\mathcal{D}}'$ show the 12 channel array of all the subbands. 

\noindent\textbf{b) DFT : }
Depending on the spatial resolution, DFT decomposes the signal into sine and cosine components. Due to the correspondence between the output of DFT and the number of input pixels, the size of the two domains is equal. Given the YCbCr image, per channel two-dimensional DFT is computed as follows:
\begin{equation}\label{2}
\begin{split}
&I_{kl}= \sum_{m=0}^{M-1}\sum_{n=0}^{N-1}a_{mn} exp^{\left\{-2\pi i\left(\frac{mk}{M}+\frac{nl}{N}\right )  \right \}} ,\\
& k=0,\dots, M-1\hspace{0.5cm}; \hspace{0.5cm} l=0,\dots ,N-1 \\
\end{split}
\end{equation}
 In our case,  M and N are 640, $a_{mn}$ is the image in the spatial domain and the exponential term is the basis function corresponding to each $kl$-th location in the fourier space. 
As shown in Figure \ref{fig:dft_comerison},  DFT  captures the nucleus density information very well.

\subsection{Feature Extraction Module (FEM)}
\label{section:feat_extraction}

Bengs \emph{et al} \cite{bengs2021multi} proposed a modified version of ResNet50 \cite{he2016deep} architecture named TResNet and experimentally proved that it works much better for multi-label classification.
  Compared to the ResNet50,  TResNet architecture contains SpaceToDepth Stem,  Anti-Alias Downsampling, In-Place Activated BatchNorm, and Novel Block-type Selection. SpaceToDepth 
design is used to minimize information loss during the max-pooled downsampling. Note that max-pooling is inherently composed of two operations: (1) evaluating the max operator densely and (2) naive subsampling. Authors in \cite{zhang2019making} introduce a low-pass filter layer to achieve Anti-`Alias Downsampling'. To make training faster, `In-Place Activated BatchNorm' is introduced for computationally efficient fusion of batch
normalization, activation layers, and memory optimization} \cite{bulo2018place}. ResNet34 uses solely `BasicBlock’ layers, which comprise two conv3$\times$3 layers as the basic building block, while ResNet50 uses
’Bottleneck’ layers, which comprise two conv1$\times$1 and
one conv3$\times$3 as the basic building block.
A better speed-accuracy trade-off
can be obtained using a `Novel Block-type Selection', which uses a mixture
of BasicBlock and Bottleneck layers. Since BasicBlock layers have a larger receptive field, they are placed at the first
two stages of the network, and Bottleneck layers at the last
two stages.  To reduce the computational cost of the Squeeze and Excite (SE) blocks,
and gain the maximal speed-accuracy benefit, they placed SE
layers only in the first three stages of the network (`Optimized squeeze') \cite{ridnik2021tresnet}.

To decouple the focusing levels of the positive and negative samples, Ridnik \emph{et al} \cite{ridnik2021asymmetric}  introduced the asymmetric loss. Given K labels, where $z_k$ in the network output, $y_k$ is the ground truth label; single  label positive class and negative class losses are formal as follows:
\begin{eqnarray}
\label{4}
\begin{split}
L_+&=\left ( 1-p \right )^{\gamma+}  +log(p),\\
L_-&=\left ( p_m \right )^{\gamma-}  -log(1-p_m),\\
\text{where}&\\
p_m&=max(p-m,0),\\
\end{split}
\end{eqnarray}
where  $L_+$ and $L_-$ are the positive class and negative class loss respectively, $p$ is the network’s output probability, $p_m$ is the shifted probability with the $(m\geq0)$ a tunable hyper-parameter   and $\gamma$ is the focusing parameter. Let ${\gamma+}$ and ${\gamma-}$ be the positive and negative focusing parameters, respectively so, to emphasize the contribution of positive samples, $\gamma- >  \gamma+$ is used. 
samples. The separate $L_+$ and $L_-$ losses help the network learn meaningful features from samples, despite their rarity.
The total loss is given as follows:
\begin{eqnarray}
\label{5}
\begin{split}
L &= yL_+ - (1-y)L_- ,\\
L_{total}&= \sum_{K=1}^{K}L(\sigma (z_k),y_k).
\end{split}
\end{eqnarray}
{$L$ represents the loss of the single-label binary classes and finally, for multi-label $L_{total}$ represents the loss for K labels binary classes with the sigmoid ($\sigma $) function. } For example: in our case, ER is a single-label with positive and negative classes, and ER, PR, HER2, etc are multi-label with individual positive and negative classes. }

We use the MS-COCO pre-trained weights for fine-tuning the TResNet-xl  \cite{ridnik2021asymmetric}. The details of the hyper-parameter setting are given in section \ref{section:implemntation} implementation details. For feature extraction, we select the best weights on behalf of mean average precision for multi-label classification on validation data and removed the last fully connected layer of the model for the features extraction module (FEM). In each, bag, the feature vectors can vary from 1 to 275.

\subsection{Malignant Region Learning} 
\label{section:att_mehanisam}
Employing FEM, we compute features of all the extracted patches. The feature vector of each patch ($640\times640$)  is of dimension $ 1 \times 2656$. 

Let a dataset: $\left\{ \left(\mathcal{ X}_{1},\mathcal{Y}_{1} \right),\left( \mathcal{X}_{2},\mathcal{Y}_{2} \right),,,\left( \mathcal{X}_{m},\mathcal{Y}_{m} \right) \right\}$ consist of a collection of samples where $\mathcal{X}$ donates an image,  $\mathcal{Y}^{(1\times k)}$  represents image labels  and $m$ donates  the number of samples. Note that $k$ represents several classes which are six in our case (ER, PR, HER2, ALN, MS, HG). After feature vector extraction, each sample corresponds to a featured bag and a set of labels. Specifically, a bag of features $\mathcal{B}$ contains a set of feature vectors $\left(\mathcal{F}_{1},\mathcal{F}_{2},\mathcal{F}_{3},,,,\mathcal{F}_{n} \right)$ where $\mathcal{F}$ is the feature vector and $n$ is the number of instances in that bag. {Note that, the value of $n$ is not fixed because of the variable number and sizes of ROIs.}  

\noindent\textbf{Attention-based MIL and Gated-MIL
} have been highly popular for the WSI classification. The neural network of attention-based MIL and Gated-MIL is shown in Figure \ref{fig:MIL_M} (a, b). The working principle of traditional attention-based MIL and Gated MIL is to pick the valuable instances in the bag and predict the bag class using those instances. However, in the embedding-based approach, all instances in the bags are used for bag prediction \cite{ilse2018attention}. Note that attention mechanism MIL and Gated-MIL were mostly used for single-factor binary or multi-factor classification problems and both used the Tanh activation function for the distinction between positive and negative instances. 

\noindent\textbf{MRL:} Inspired by \cite{zaigham2022clustering}, we propose an MRL attention mechanism. However, as compared to \cite{zaigham2022clustering} who use the softmax activation function without the dropout layer. { In our framework, the instances of a bag belong to malignant regions of a WSI, so all instances are equally important. To maintain the similar importance of all instances in a bag, we use the two-stream attention mechanism to automatically discover the parameters and to improve the instance similarity of the beg.  In the first stream, we use the softplus activation function with dropout (s1), and in the second stream, we employ the sigmoid activation function with dropout (s2). We use the Softplus activation function because  it is approximately linear for \(x\epsilon [0,1]\) and further we divide the product of (s1,s2) 
with the sum of the (s1,s2) to maintain instance similarity in the neural network of the MRL mechanism shown in Figure \ref{fig:MIL_M} (c). 
Logically, the range of Tanh is (-1 to 1) so the range of MIL and Gate-MIL attention mechanism is (-1 to 1). But in our scenario, we require close linear values, so we use the two streams and further mathematically process them for more close-range values to improve the bag instance's similarity.  Finally, because of the different sizes of bags, we use the permutation-invariant mean operators for the weighted average aggregation of the bag instances. }


\begin{figure}[!ht]
\centering
\includegraphics[width=0.5\textwidth]{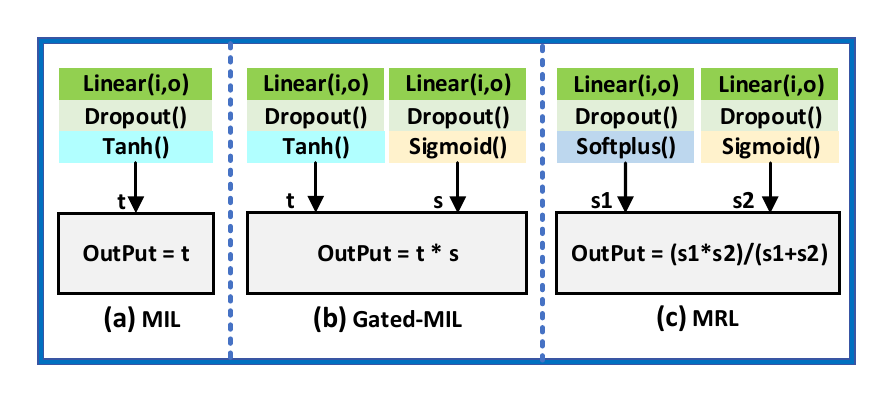}
\caption{Neural Network Architecture of MIL, Gated-MIL, and MRL attention mechanism. 
The output range of the (a) MIL and (b) Gated-MIL is from (-1 to 1) but the range of (c) MRL mechanism exists between (0 to 1). }
\label{fig:MIL_M}
\end{figure}


 \begin{table*}[cols=12,pos=!ht]\centering
 \caption{The details of the Breast Cancer Core Needle Biopsy (BNCB) dataset: The first three factors have binary and the last three factors have multi-classes. The class distribution of the Molecular Subtype is balanced but all the other factors have imbalanced classes.}
\label{Class details}
\small
\begin{tabular}{cc|cc|cc|cc|cc|cc}
\toprule
\multicolumn{2}{  c |}{\textbf{ER}} & \multicolumn{2}{ c |}{\textbf{PR }} & \multicolumn{2}{ c |}{\textbf{HER2}} & \multicolumn{2}{ c |}{\textbf{Histological Grading}} & \multicolumn{2}{ c |}{\textbf{Molecular Subtype}} & \multicolumn{2}{ c }{\textbf{ALN Status}} \\
\midrule 
\centering
Positive & 831 & Positive & 790 & Positive & 277 & G1 & 38 & Luminal A & 288 & N0 & 655 \\ 
Negative & 227 & Negative & 268 & Negative & 781 & G2 & 518 & Luminal B & 372 & N+($>$2) & 193 \\ 
 & & & &&&G3 & 370 & Triple Negative & 125 & N+(1-2) & 210 \\ 
& & & & &&&&HER2(+) & 273 & & \\ 
\bottomrule
\end{tabular}
\end{table*}

\begin{table*}[!h]\centering
\caption{The details of the Breast Cancer Core Needle Biopsy (BNCB) dataset after multi-classes to binary classes conversion with the class difference:  The class distribution of Histological grading and ALN status looks balanced but Molecular subtype has the worst imbalanced classes with the other ER, PR, and HER2 labels.}
\label{new classes}
\small
\scalebox{0.91}{
\begin{tabular}{cc|cc|cc|cc|cc|cc}
\toprule
\multicolumn{2}{  c |}{\textbf{ER}} & \multicolumn{2}{ c |}{\textbf{PR }} & \multicolumn{2}{ c |}{\textbf{HER2}} & \multicolumn{2}{ c |}{\textbf{Histological Grading}} & \multicolumn{2}{ c |}{\textbf{Molecular Subtype}} & \multicolumn{2}{ c }{\textbf{ALN Status}} \\
\midrule 
\centering
Negative & 227 & Negative & 268 & Negative & 781 & well differentiated & 556 & Any Positive & 933 & N0 & 655 \\ 
Positive & 831 & Positive & 790 & Positive & 277 & Poorly differentiated & 370 & Triple-Negative & 125 & N+ & 403 \\ 
Difference & 526 & Difference & 456 & Difference & 408 & Difference & 186 & Difference & 808 & Difference & 252\\
\bottomrule
\end{tabular}}
\end{table*}
\section{ Experiments and Results } 
\label{section_ER}
In this section, we discuss the datasets used, our experimental setups, implementation details, evaluation metrics, and results for multi-factors on the BNCB dataset and binary-class on the BACH  dataset.

\subsection{Datasets}
\label{sec:formatting}
To thoroughly evaluate our proposed approach, we have used recent  large-scale publicly available datasets: Breast Cancer Core Needle Biopsy (BCNB) \cite{xu2021predicting} and Breast cancer histology images dataset (BACH) \cite{aresta2019bach}


\noindent\textbf{Breast Cancer Core Needle Biopsy (BCNB): }The dataset contains the  H\&E stained tissues WSI's of core needle biopsy from early breast cancer patients with clinical information and splits cohorts. Clinical information is also based on six types of labels with ER, PR, HER2, MS, HG, and ALN status. The train, test, and validation cohorts of the BNCB dataset are 6:2:2. There are 1058 WSI's with a size of 11486 $\times$ 31000 pixels, and malignant regions are annotated by pathologists. Factor details of each label are shown in Table \ref{Class details}. 

Histological grading, molecular sub-type, and ALN status consist of multiple labels. Table \ref{Class details} shows that the histological grading is based on three labels G1, G2, and G3, and they are distinguished based on cell structure. The molecular subtype is of four types: Luminal A, Luminal B, triple negative,  and HER2 (+), similarly, ALN status is based on three levels: N0, N+(>2), and N+(1-2). 

For the binary classes, we convert the histological grading classes G1 and G2 into  `well differentiated' and G3 to `poorly differentiated'  \cite{rakha2010breast} labels.  Similarly, for molecular sub-types Luminal A, Luminal B, and HER2(+); the label is negative if all of these are negative (triple-negative) and positive otherwise \cite{ma2019breast}. For ALN status, N0 is labeled as negative and the other two levels (N+(>2) and N+(1-2)) are labeled as positive \cite{xu2021predicting}. The dataset details are shown in Table \ref{new classes} with the class difference. Note that the class difference varies from 186 to 526 (the maximum in MS is 808 and the minimum in HG is 186).
\begin{table*}[b!]
\caption{TResNet-xl Results BNCB dataset: Comparison of the spatial and frequency domains. Frequency (DWT) and Spatial (RGB) performance on the test and validation is reciprocal. The frequency domain (DFT) outperforms both cohorts.  }
\label{CNN_results}
 \centering
 \begin{tabular}{lllllllll}
 \toprule

\textbf{Domain} & \textbf{mAP} & \textbf{AUC} & \textbf{ACC\%} & \textbf{SPEC\%} & \textbf{SEN \%} & \textbf{PPV\% }& \textbf{NPV\% }& \textbf{F1\% } \\ \midrule 
\multicolumn{9}{ c }{\textbf{Train Results}} \\ 
\midrule 

  Spatial (RGB) & 99.27 & 0.99 & 96.52 & 95.52 & 99.20 & 93.00 & 98.77 & 95.76 \\ \hline
  Frequency (DFT) & 99.99 & 1.00 & 99.94 & 99.74 & 100.00 & 99.92 & 100.00 & 99.96 \\ \hline
  Frequency (DWT) & 91.27 & 0.95 & 90.60 & 86.74 & 84.62 & 85.25 & 93.22 & 84.53 \\ \midrule 
\multicolumn{9}{ c }{\textbf{Validation Results}} \\ \midrule
 Spatial (RGB) & \underline{65.47} & \underline{0.75} & 73.88 & 54.01 & \textbf{68.32} & \underline{57.99} & \textbf{82.57} & \underline{60.59} \\ 
 Frequency (DFT) & \textbf{68.62} & \textbf{0.78} & \underline{76.90} & \textbf{65.47} & \underline{65.02} & \textbf{63.61} & \underline{76.99} & \textbf{63.57} \\ 
 Frequency (DWT) & 60.59 & 0.71 & \textbf{78.42} & \underline{59.07} & 55.99 & 52.29 & 70.70 & 53.27 \\ 
  \midrule

\multicolumn{9}{ c }{\textbf{Test Results}} \\\midrule
 Spatial (RGB) & 62.77 & 0.75 & 73.87 & 56.22 & \textbf{67.23} & 58.58 & 77.80 & \underline{61.33} \\ 
 Frequency (DFT) & \textbf{67.84} & \textbf{0.79} & \textbf{77.75} & \textbf{70.14} & \underline{66.83} & \textbf{63.36} & \textbf{79.04} & \textbf{64.95} \\ 
 Frequency (DWT) & \underline{65.02} & \underline{0.76} & \underline{75.95} & \underline{64.02} & 62.51 & \underline{62.03} & \underline{78.18} & 61.19 \\ 

\bottomrule
 \end{tabular}

\end{table*}
In our experiments, we use the annotations provided along with the dataset for ROI extraction and employ both overlapping and non-overlapping patches. However, we observed that the performance on overlapped patches is much better than the separate patches. The BNCB dataset contains 1058 images in total, however, we exclude some of the very low image quality and poorly annotations patches.

\noindent\textbf{{BACH: Grand challenge on breast cancer histology images dataset} :}  
The BACH dataset's  \cite{aresta2019bach} first release  (Train set) contains four hundred  H\&E stained microscopy images of BC patients with image-level labels. The size of each image is ($2048 \times 1536$) and each image is labeled as normal, benign, in situ carcinoma, and 
invasive carcinoma. For a fair comparison, we split the dataset by following \cite{bhowal2021fuzzy}, further we use the 4 fold (A, B, C, D) cross-validation with the ratio of 300:100,   the dataset split detail shown in Table \ref{bach_dataset_detail}. Following \cite{rakhlin2018deep,li2023manifold}, we use the same methods for  2-class non-carcinomas (normal and benign) vs. carcinomas (in situ and invasive) classes distribution. For data augmentation, patches, and feature extraction, we follow the \cite{rakhlin2018deep}.

\begin{table}[width=.9\linewidth,cols=4,pos=h!]
\caption{BACH: Grand challenge on breast cancer histology images dataset split detail }\label{bach_dataset_detail}
\begin{tabular*}{\tblwidth}{@{} LLLL@{} }
\toprule
 \textbf{Cohort} & \textbf{Non-carcinomas } & \textbf 
 {Carcinomas} &\textbf{Total}
  \\ 
\midrule 
 Train & 150 & 150 &300\\ 
 Test& 50 & 50 &100\\

\bottomrule
\end{tabular*}
\end{table}

\subsection{Experimental Results BNCB}
\label{section:implemntation}
\noindent\textbf{Implementation details:} We train the TResNet on spatial and frequency domains for feature extraction and train the MRL module on these features for factors prediction. For both model training, we use the asymmetric loss function to account for class imbalance i.e., for asymmetric
focusing, we set the $\gamma+ = 0$ and $\gamma- = 1$. As an optimizer, Adam \cite{kingma2014adam} is employed with a learning rate of 0.001, weight decay of 0.001, and clip rate of 0.08. The feature extraction module (FEM) is trained for around 30 epochs with a batch size of 16 and MRL training is done for 25 epochs with a batch size of 1. We use a batch size of 1 because the size of the bags is variable with some bags containing only one feature instance because of small malignant regions. For FEM training and feature extraction, we use the zero mean and 1-std normalization according to the domain (spatial or frequency) channels.

\noindent\textbf{Evaluation Metrics:} 
To evaluate the performance of the proposed approach,  Area under the Curve (AUC), accuracy (ACC), sensitivity (SENS), specificity (SPEC), the positive predictive value (PPV), and negative predictive value (NPV) \cite{xu2021predicting}. For better evaluation, we also used the mean Average Precision (mAP) for all factors and the F1 score for individual factors.

\begin{table*}[!Hb]
\caption{Individual factors test data results: Without MRL results in comparison of spatial (RGB) and frequency (DFT) domains. All metrics showing the performance of the DFT domain is much better than RGB. }
 \label{CNN_results_detail}
\scalebox{1}{
 \small
 \centering
 \begin{tabular}{lllllllll}
 \toprule
 \textbf{Domain} & \textbf{factor } & \textbf{AUC} & \textbf{Acc\%} & \textbf{Spec\%} &
 \textbf{Sen\%} & \textbf{PPV\%} & \textbf{NPV\%} & \textbf{F1\%} \\ \midrule 
 DFT & ER & \textbf{0.85} & \textbf{81.62} & \textbf{46.67} & 92.86 & \textbf{84.48} & \textbf{67.74} & \textbf{88.44} \\ 
 RGB & ER & 0.80 & 77.84 & 17.78 & \textbf{97.14} & 78.61 & 66.67 & 86.90 \\ \midrule
 DFT & PR & \textbf{0.80} & \textbf{79.46} & \textbf{41.51} & 94.70 & \textbf{80.13} & \textbf{75.86} & \textbf{86.81} \\ 
 RGB & PR & 0.74 & 74.05 & 15.09 & \textbf{97.73} & 74.14 & 72.73 & 84.31 \\ \midrule
 DFT & HER2 & \textbf{0.79} & 75.14 & 82.44 & \textbf{57.41} & 57.41 & \textbf{82.44} & \textbf{57.41} \\ 
 RGB & HER2 & \textbf{0.79} & \textbf{76.22} & \textbf{87.02} & 50.00 & \textbf{61.36} & 80.85 & 55.10 \\ \midrule
 DFT & ALN & \textbf{0.83} & \textbf{76.22} & \textbf{81.03} & 68.12 & \textbf{68.12} & 81.03 & \textbf{68.12} \\ 
 RGB & ALN & 0.79 & 68.65 & 62.07 & \textbf{79.71} & 55.56 & \textbf{83.72} & 65.48 \\ \midrule
 DFT & MS & \textbf{0.75} & 83.24 & 90.18 & \textbf{31.82} & 30.43 & \textbf{90.74} & \textbf{31.11} \\ 
 RGB & MS & 0.71 & \textbf{86.49} & \textbf{95.71} & 18.18 & \textbf{36.36} & 89.66 & 24.24 \\\midrule 
 DFT & HG & \textbf{0.75} & \textbf{70.81} & \textbf{78.99} & 56.06 & \textbf{59.68} & \textbf{76.42} & \textbf{57.81} \\ 
 RGB & HG & 0.67 & 60.00 & 59.66 & \textbf{60.61} & 45.45 & 73.20 & 51.95 \\ \bottomrule
 
 \end{tabular} }
\end{table*}

\subsubsection{Results without MRL}
\label{section:FEM}
    We evaluate our model without attention networks i.e., using only TResNet CNN. Specifically, we took the mean aggregation of TResNet features before the final prediction of all the patches within an image and predicted the BC factors. The experimental results for TransNet features on testing and validation data of the BNCB dataset for spatial and frequency domain (both DFT and DWT) are shown in Table \ref{CNN_results}. The results on several different metrics show that the frequency domain (DFT) performs much better as compared spatial domain.  Due to the imbalance dataset (see Table \ref{Class details} ), we also evaluate and demonstrate the performance of spatial (RGB) and frequency (DFT) domains based on the individual factor in  Table \ref{CNN_results_detail}. 

For diagnostic purposes, it's desirable to have a factor diagnosis that is both highly sensitive and highly specific. As compared to the spatial domain (RGB),  the values of the frequency domain (DFT) show a better correlation between sensitivity and specificity {by reducing the false negative predictions}. In addition, as shown in {Table \ref{CNN_results} shows}, better AUC and F1-score results also demonstrate that the performance of the frequency domain (DFT) is much better than the spatial domain.  \\
Table \ref{CNN_results_detail} shows the performance of individual factors and is explained as follows.
The overall performance indicators AUC and F1 show that the frequency domain (DFT) outperforms.  The sensitivity of  spatial domain (RGB) for \noindent\textbf{ER}, \noindent\textbf{PR} and \noindent\textbf{HG} indicate  4.28\%,  2.31\%, and 4.55\% respectively higher than the frequency domain (DFT). That shows the spatial domain is also useful and contains complementary information.
     
\noindent\textbf{HER2 and ALN:} a trade-off between the specificity and sensitivity evaluation parameters shows that the frequency domain (DFT) holds some false negative results and the spatial domain  (RGB) holds some false positive results. AUC is the same for HER2  but the F1 of the frequency domain (DFT) is  2.31\% higher than the spatial domain (RGB) and for ALN, the  AUC, and F1 of the frequency domain (DFT) are  4\%, and 2.64\% respectively higher than the spatial domain (RGB). But the difference between the specificity and sensitivity of the frequency domain is 12.91 and in the spatial domain is 17.63 which shows the frequency domain (DFT) is less biased toward the dominant class.

\noindent\textbf{MS:} The AUC, F1, and other evaluation matrices show a similar pattern of experimental results.
\begin{table}
\caption{TResNet-xl test data results without MRL on different patch sizes using the spatial domain (RGB)}\label{patches_results_auc_nonoverlapped}
\begin{tabular*}{\tblwidth}{@{} LLLLLL@{} }
\toprule
 \textbf{Patches size} & \textbf{mAP } & \textbf{AUC} & \textbf{Acc\%} & \textbf{Spec\%} &
 \textbf{Sen\%} \\ 
\midrule 
 224 $\times$ 224 & 55.68 & 0.65& 71.50 & 62.07 & 58.10\\ 
 448 $\times$ 448 & 55.80&0.65&71.90&62.76&57.04 \\ 
 640 $\times$ 640 & \textbf{59.90}&\textbf{0.71}&\textbf{74.15}&\textbf{63.10}&\textbf{61.67} \\ 
  
\bottomrule
\end{tabular*}
\end{table}
\begin{table*}[!hb]
\caption{Malignant Region Learning Results: Comparison of our proposed method on the BNCB dataset by using the different combinations of input domains with the different stream models. Best scores are shown in bold, and second, best are underlined. }

\label{MS:AIL_Rsults}
\small
 
 \scalebox{0.97} {
 \begin{tabular}{l|lllllllll}
 \toprule

 
 \textbf{Domain} & \textbf{Model} & \textbf{mAP} & \textbf{AUC} & \textbf{ACC\%} & \textbf{SPEC\%} & \textbf{SEN \%} & \textbf{PPV\% }& \textbf{NPV\% }& \textbf{F1\% } \\ 
\midrule 
\multicolumn{10}{ c }{\textbf{Validation Results}} \\ \midrule

 Spatial(RGB) &SS-MRL & 64.09 & 0.77 & 75.73 & 66.97 & 67.17 & 61.34 & 76.43 & 63.71 \\ 
 Frequency(DWT) & SS-MRL & 67.25 & 0.77 & 77.10 & \textbf{72.41} & 61.32 & 62.73 & 74.57 & 61.58 \\ 
 Frequency(DFT) &SS-MRL & 67.36 & 0.77 & 77.00 & 69.22 & \underline{67.40} & 63.96 & 75.48 & 65.22 \\ 
 Spatial(RGB), Frequency(DWT) & JS-MRL & 67.64 & \underline{0.78} & 77.00 & 72.12 & 62.08 & 62.62 & 75.70 & 61.94 \\ 
 Spatial(RGB), Frequency(DFT) & JS-MRL & 68.17 & \textbf{0.79} & 76.61 & 71.77 & \textbf{67.59} & \underline{64.59} & 73.90 & \textbf{65.58} \\ 
 Frequency(DWT,DFT) & JS-MRL & \underline{69.19} & \underline{0.78} & \underline{77.49} & 71.04 & 65.26 & 63.64 & \underline{77.29} & 63.88 \\
  Spatial(RGB), Frequency(DWT,DFT) &JS-MRL & \textbf{69.63} & \textbf{0.79} & \textbf{78.36} & \underline{72.21} & 66.71 & \textbf{65.21} & \textbf{79.05} & \underline{65.41} \\ 
\toprule

\multicolumn{10}{ c }{\textbf{Test Results}} \\ \midrule

 Spatial(RGB) & SS-MRL & 62.90 & 0.73 & 73.49 & 65.89 & \underline{64.51} & 59.05 & 73.52 & 61.30 \\ 
 Frequency(DWT) & SS-MRL & 66.99 & 0.73 & 76.51 & \textbf{72.51} & 63.35 & \textbf{65.48} & 75.91 & 62.96 \\ 
 Frequency(DFT) & SS-MRL & 66.92 & \underline{0.76} & 75.34 & 68.84 & 63.77 & 61.91 & 75.23 & 62.12 \\ 
 Spatial(RGB), Frequency(DWT) & JS-MRL & 66.63 & \underline{0.76} & 75.93 & \underline{72.38} & 63.50 & 63.47 & 75.03 & 62.94 \\ 
 Spatial(RGB), Frequency(DFT) &JS-MRL & 67.79 & 0.74 & 76.22 & 71.51 & 62.78 & 63.98 & 76.26 & 62.39 \\ 
 Frequency(DWT,DFT) & JS-MRL & \textbf{69.81} & \underline{0.76} & \underline{76.90} & 71.41 & 64.09 & 64.29 & \underline{77.94} & \underline{63.04} \\ 
 
 Spatial(RGB), Frequency(DWT,DFT) &JS-MRL & \underline{69.16} & \textbf{0.78} & \textbf{77.57} & 72.03 & \textbf{65.13} & \underline{64.72} & \textbf{78.35} & \textbf{63.81} \\ 
 
\bottomrule
 \end{tabular} }
\end{table*}

\noindent\textbf{Evaluation using different patch sizes}
Table \ref{patches_results_auc_nonoverlapped} shows the model performance using different patch sizes using RGB data.  Specifically, we evaluate on patch size of 224 $\times$ 224, 448 $\times$ 448, and 640 $\times$ 640. 
Since large-sized ($640 \times 640$) patches hold better spatial relationships and produce improve classification results,  we use the patch size of $640 \times 640$ in our experiments.

\subsubsection{Results with MRL} 
To evaluate the performance of our joint stream malignant region learning framework, we perform a range of experiments. 


\begin{table*}[h]\centering
\caption{Attention Mechanism Results: Performance comparison of MRL with the state-of-the-art MIL and Gated-MIL (G-MIL) \cite{ilse2018attention} in our proposed JS:MRL architecture. Best scores are shown in bold, and second, best are underlined.}
\label{MIL_Results}

\scalebox{0.95}{
 \centering
 \begin{tabular}{llllllllll}
  \toprule
\multicolumn{10}{ c }{\textbf{Test Results}} \\ 
\midrule 

 \textbf{Mechanism} & \textbf{Domain} & \textbf{mAP} & \textbf{AUC} & \textbf{ACC\%} & \textbf{SPEC\%} & \textbf{SEN \%} & \textbf{PPV\% }& \textbf{NPV\% }& \textbf{F1\% } \\ 
 
\midrule 
MIL &Spatial(RGB)&61.39&0.72   &72.61&63.01&\textbf{68.76}&57.18&74.82&62.19\\
G-MIL&Spatial(RGB)&59.39&0.70&73.15&64.54&65.55&57.31&74.87&61.07\\
MIL &Spatial(RGB), Frequency(DWT,DFT) & \underline{68.76} & \underline{0.74} & 77.39 & 70.72 & \underline{65.16} & 63.60 & \textbf{78.74} & \underline{63.67} \\ 
 G-MIL &Spatial(RGB), Frequency(DWT,DFT) & 67.96 & 0.71 & \textbf{77.75} & \underline{71.93} & 63.53 & \textbf{65.84} & 77.72 & 63.35 \\ 

 MRL & Spatial(RGB), Frequency(DWT,DFT) & \textbf{69.16} & \textbf{0.78} & \underline{77.57} & \textbf{72.03} & 65.13 & \underline{64.72} & \underline{78.35} & \textbf{63.81} \\ 
  
\bottomrule
 \end{tabular}}
 \end{table*}

The validation and test performances for different modalities are shown in Table \ref{MS:AIL_Rsults}. The best performance is where the image bag of two domains (RGB and frequency) is concatenated for a single training set. The features of spatial (RGB) and frequency (DWT) domains helped the DFT to improve the performance of the model. As compared to MRL trained only on the spatial domain, the MRL trained on both spatial and frequency domains performs much better. 
\begin{figure}[!h]
\centering
\includegraphics[width=0.5\textwidth]{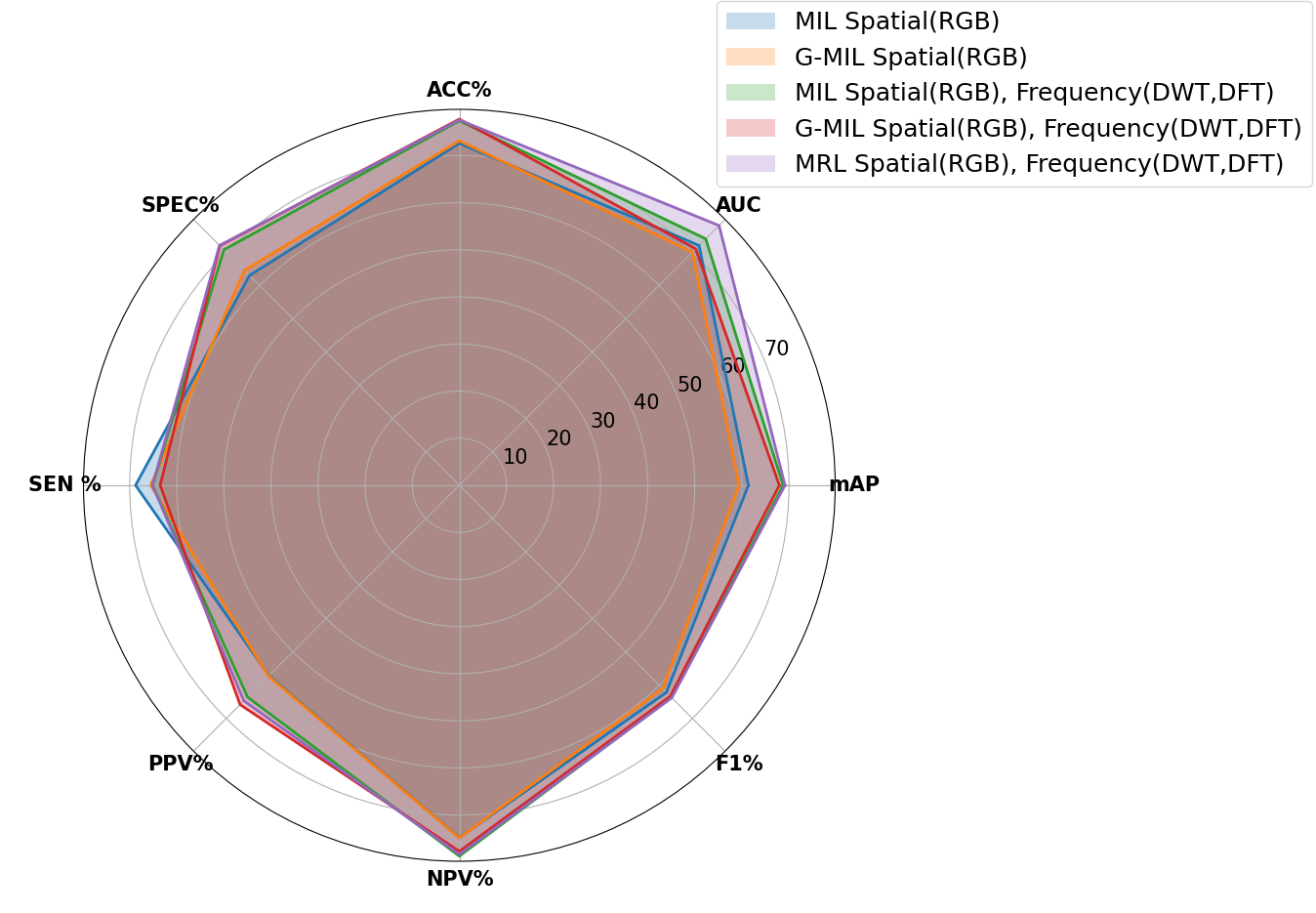}
\caption{Graphical representation comparing the results of the proposed attention mechanism against state-of-the-art methods.}
\label{fig:Spider_chart}
\end{figure}
 
For evaluating the performance of the proposed MRL mechanism, we use the state-of-the-art attention mechanisms  MIL and Gated-MIL \cite{ilse2018attention} in our proposed architecture. The performance of our proposed MRL mechanism is much better than the previously used state-of-the-art approach. 
The results of our experiments are shown in Table \ref{MS:AIL_Rsults}. The performance of the frequency domain is better than the spatial domain using the single-stream malignant region learning (SS: MRL) and the performance of joint stream malignant region learning using spatial and DFT features information, is much better than the SS: MRL. Finally, joint stream malignant region learning using  spatial, DWT and DFT features information outperforms the difference in mAP
(6.26\%), AUC (5\%), ACC (4.08\%), SPEC (6.14\%), SEN  (2.2\%), PPV (5.67\%), NPV (4.83\%), and F1 (2.51\%), then the SS:MRL spatial domain.  
 The AUC scores with receiver operating characteristic (ROC) curves of the individual factors are shown in Figure \ref{fig:three graphs}. { The AUC score for all factors indicates that the MRL performance is better.  {The AUC score of MRL for ER, PR, HER2, MS, and HG  is higher than the MIL and Gated MIL and for ALN  the performance of MRL is similar to MIL and better than the Gated MIL}. 


\begin{table*}[h]

\caption{Performance comparison in prediction of ALN status (N0 vs. N(+)) with the baseline method DL-CNB, DL-CNB-C \cite{xu2021predicting}  and Dsnet \cite{xiang2022dsnet}.   }
\label{tab:SOTA}
\centering
\scalebox{1}{

 \begin{tabular}{|l|l|l|l|l|l|l|l|}
 \hline
 \textbf{Label } & \textbf{AUC} & \textbf{Acc\%} & \textbf{Spec\%} &
 \textbf{Sen\%} & \textbf{PPV\%} & \textbf{NPV\%} & \textbf{F1\%} \\ \hline
  
        DL-CNB \cite{xu2021predicting} & 0.82&74.77&70.90&80.95&63.55&85.69&-\\ \hline
        DL-CNB-C \cite{xu2021predicting} & \textbf{0.83}&\textbf{75.69}&67.16&\textbf{89.29}&63.03&\textbf{90.9}1&-\\ \hline
        Dsnet \cite{xiang2022dsnet} & 0.80&-&-&-&-&-&-\\ \hline
        Our JS-MRL & 0.82 & 74.59 & \textbf{77.59} & 69.57 & \textbf{64.86} & 81.08 & 67.13  \\ \hline
 \end{tabular} }
\end{table*}

 \textbf{Comparison with the state-of-the-art: }{ Table \ref{MIL_Results} shows that the performance of the state-of-the-art MIL and Gated using the joint stream (spatial, DWT, DFT) is much better than the single-stream spatial domain. Furthermore, the visual illustration is shown in Figure \ref{fig:Spider_chart}
The overall performance of  MIL (spatial, DWT, DFT)  in mAP (7.37\%), AUC (2\%), and F1 (1.48\%) is higher than the MIL (spatial), similarly, the gated mil (spatial, DWT, DFT) is better than the gated-mil (spatial) with the difference of mAP (8.57\%), AUC (1\%) and F1 (2.28\%).
Finally,  the joint
stream malignant region learning (JS: MRL) from spatial, DWT, and DFT outperforms the difference in mAP ( 7.77\%), AUC (6\%), and F1 (1.62\%) than the MIL single stream spatial domain. }\\
\begin{figure*}[h]
 \centering
 \begin{subfigure}[b]{0.32\textwidth}
 \centering
 \includegraphics[width=\textwidth]{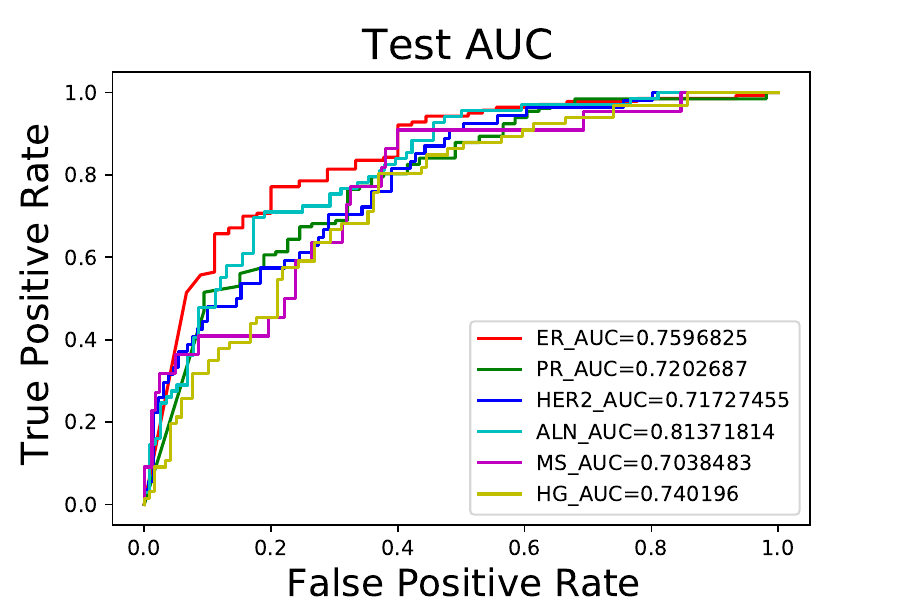}
 \caption{MIL}
 \label{fig:}
 \end{subfigure}
 \hfill
 \begin{subfigure}[b]{0.32\textwidth}
 \centering
 \includegraphics[width=\textwidth]{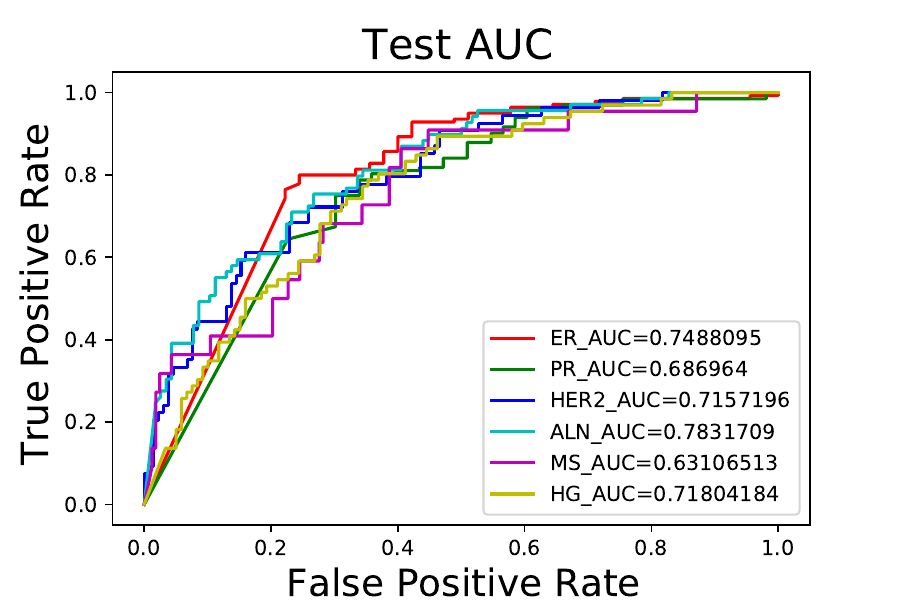}
 \caption{Gated-MIL}
 \label{fig:three sin x}
 \end{subfigure}
 \hfill
 \begin{subfigure}[b]{0.32\textwidth}
 \centering
 \includegraphics[width=\textwidth]{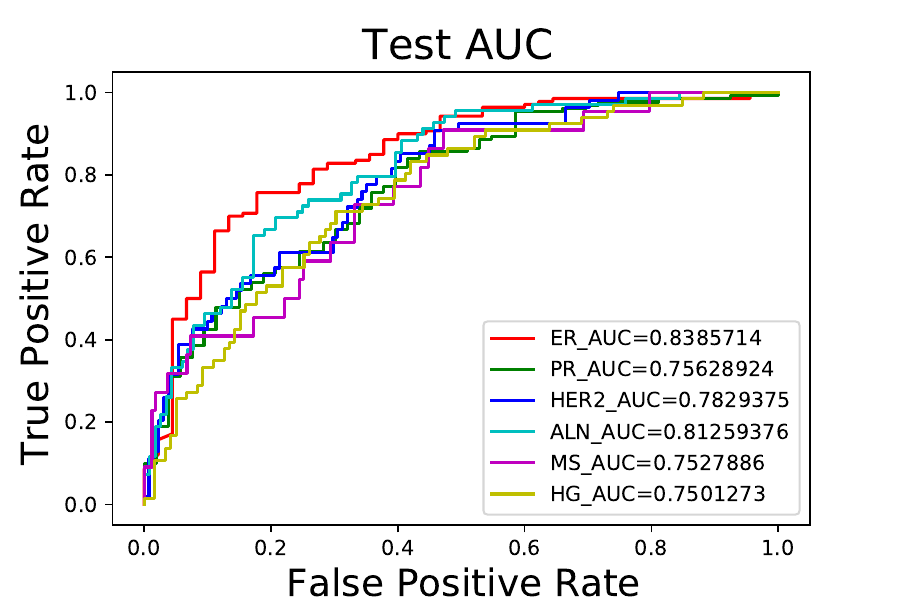}
 \caption{MRL}
 \label{fig:five over x}
 \end{subfigure}
 \caption{Receiver operating characteristic (ROC) curves with Area under the curve (AUC) values show the individual label (ER, PR, HER, ALN, MS, HG) performance comparison between the state-of-the-art MIL, Gated-MIL, and MRL in our proposed JS: MRL. The results indicate that the performance of our MRL is much better than the state-of-the-art mechanisms.}
 \label{fig:three graphs}
\end{figure*}
\noindent\textbf{Comparison with the state-of-the-art for ALN status: } The results in Table \ref{tab:SOTA}  show that JS: MRL benefits the multiple domains for training, and therefore obtaining higher performance compared to a single spatial domain and a baseline DL-CNB \cite{xu2021predicting} and Dsnet \cite{xiang2022dsnet}. A MIL-based \cite{ilse2018attention} model produced only a single-factor prediction. Using the same dataset, DL-CNB, DL-CNB -C \cite{xu2021predicting}  and Dsnet \cite{xiang2022dsnet}  achieved the 0.82, 0.83, and 0.80     AUC scores respectively by taking into account ALN status only.  On the contrary, we achieved a better than Desnt and similar DL-CNB AUC score of 0.82 including five other factors ER (0.84), PR (0.76), HER2 (0.78), MS (0.75), and HG (0.75) in addition to ALN status.


\subsection{Experimental Results of BACH}
\noindent\textbf{Implementation detail:} Following \cite{rakhlin2018deep}, we extract the patches from microscopy images of size $2048 \times 1536$ and use ImageNet dataset pre-trained weights from ResNet-50 \cite{russakovsky2015imagenet} for feature extraction. We compare the performance of our proposed MRL mechanism with the MIL and Gated-MIL mechanism for the single-label binary class problem. For {training}, we follow the section \ref{section:implemntation}  implementation details for the BNCB. Specifically, we use the single-label asymmetric focusing loss \cite{ridnik2021asymmetric}. 
We set the $\gamma+ = 0$ and $\gamma- = 0$. Adam \cite{kingma2014adam} is employed with a learning rate of 0.001 and weight decay of 0.001. The evaluation is done using accuracy (Acc), True positive (TP), true Negative (TN), False Positive (FP), and False negative (FN).

\subsubsection{BACH Results}
We evaluate the binary-class classification performance of our proposed MRL mechanism. Table \ref{BACH results} shows the differentiable performance of the MRL mechanism with MIL and Gated MIL \cite{ilse2018attention} on the BACH dataset. The predicted results show the performance of MRL is better than the MIL and Gated-MIL for all folds. The 4-fold cross-validation average accuracy was 87\%. Out of 50 carcinomas cases, on average  6 to 7 cases were missed and out of 50 non-carcinomas cases, almost 5 cases were missed. 

\begin{table}[width=1\linewidth,cols=7,pos=h]
\centering
\caption{Table shows the comparison  of our proposed MRL and state-of-the-art MIL  \cite{ilse2018attention} . The table shows the  performance of our MRL is better than the MIL}
\label{BACH results}
\begin{tabular}{p{1.1cm}|p{1cm}p{0.5cm}p{0.5cm}p{0.5cm}p{0.5cm}p{1.2cm}} 
\toprule
 \textbf{Method} & \textbf{Matrices} & \textbf{A} & \textbf{B} & \textbf{C} & \textbf{D}& \textbf{Average} \\ 
\midrule
\multirow{5}{*}{MIL\cite{ilse2018attention}}
& Accuracy  & 87\% & 85\%  & \textbf{85}\%  & 91\%  & 87\%  \\ 
& TP & \textbf{45} & 43 & \textbf{40} & \textbf{50} & \textbf{44.5} \\
& TN & 42 & 42 & \textbf{45} & 41 & 42.5 \\
& FP & 5 & 7 & 10 & 0 & 5.5 \\ 
& FN & 8 & 8 & 5 & 9 & 8.5 \\ 
\midrule
\multirow{5}{*}{GMIL\cite{ilse2018attention}}
& Accuracy  & 86\% &\textbf{86}\%  & 82\%  & 87\%  & 85.25\%  \\ 
    & TP & 44 & \textbf{44} & 39 & 44 & 42.75 \\
& TN & 42 & 42 & 43 & 43 & 42.5 \\
& FP & 6 & 6 & 11 &6 & 7.25 \\ 
& FN & 8 & 8 & 7 & 7 & 7.50 \\ 
\midrule
\multirow{5}{*}{MRL}
& Accuracy  & \textbf{88}\% & 85\%  & \textbf{85}\%  & \textbf{95}\%  &\textbf{88.25} \%  \\ 
& TP & 41 & 42 &\textbf{40}  &\textbf{50}  &43.25  \\
& TN & \textbf{47} & \textbf{43} & \textbf{45} &\textbf{45}  &\textbf{45}  \\
& FP & 9 & 8 & 10 &0  &6.75  \\ 
& FN & 3 & 7 & 5 & 5 &5  \\ 
\midrule
\end{tabular}
\end{table}
\section{Discussion}
\label{section_discusion}
We demonstrate ER, PR, HER2, ALN status, MS, and HG status estimation using a machine-assisted model trained on a publicly available BNCB dataset of H\&E stained tissue images with malignant region annotation. The BNCB dataset is also challenging due to irregular shapes, variable size, several equally important dissimilar malignant regions, and imbalanced factor classes.  Concatenated feature bags from multiple domains allow the model to focus on different insights and combining both spatial and frequency information leads to prediction improvement. With a multi-domain instance bag as input and multiple predictions as output, the models produce attention focused on non-similar instances of the bags. This fact could be explained considering the factoid that multi-domain input bags and the attention mechanism of JS:MRL allow the model to have feature representation including spatial and frequency information from different domains. The results obtained show that JS:MRL benefits the multiple domains for training and optimization of the network weights, obtaining higher performance compared with a single spatial domain and a baseline DL-CNB \cite{xu2021predicting}. \\ Previously, by using the same dataset, its distribution, and the feature vectors of 10 patches (256×256) pixel per bag, DL-CNB achieved the 0.81 AUC score by taking into account ALN status only. On the contrary, we achieved a similar AUC score of 0.81 by five other factors ER (0.84), PR (0.76), HER2 (0.78), MS (0.75), and HG (0.75) in addition to ALN status by using the different number of feature vectors of patchs (640 × 640) per bag. Determining the status of IHC-derived molecular markers and comparing them with H\&E stained tissues, could reduce variability in predictions. In addition, by using digital workflows, the initiation of treatment might be expedited resulting in improved clinical outcomes. In this work, we lay a foundation for future studies to compare the performance of a pathologist’s clinical workflow with and without machine-assisted solutions. Generally, our study represents an enhancement of machine-assisted pathology and demonstrates machine-assisted pathology’s potential to improve breast cancer prognosis.
Our study aimed to predict several important factors (ER, PR, HER2, ALN status, MS, HG status) crucial for breast cancer diagnosis and prognosis using H\&E stained Whole Slide Images (WSI). The results demonstrate the significant performance of our proposed method. However, it is important to note that our approach heavily relies on having annotated whole slide images initially, which could limit its applicability without sufficient annotated data.

\section{Conclusions}
\label{section_conclusion}
In this paper, we propose a machine-assisted framework for the classification of multiple factors including  Estrogen receptor (ER), Progesterone receptor (PR), Human epidermal growth factor receptor 2 (HER2) gene, Histological grade (HG), Auxiliary lymph node (ALN) status, and molecular subtype by using only H\&E stained WSI of core needle biopsy Histopathological breast tissue images. In the framework, we introduce the malignant region learning mechanism and compare it with the state-of-the-art MIL and Gated-MIL mechanisms. We show that combining features from spatial and frequency domains help to get improved performance. One of the limitations of our proposed method is that it is dependent on the annotated whole slide images. In the future, we are keen to explore weakly or unsupervised learning approaches to reduce dependency on annotated data and extend our method to analyze H\&E stained WSIs from other organs.

\bibliographystyle{cas-model2-names}
\bibliography{revised}

\end{document}